**ARTICLE TYPE**

# Facetted patchy particles through entropy-driven patterning of mixed ligand SAMS

Aaron Santos,[a,b] Jaime Andres Millan,[c] and Sharon C. Glotzer*[a,c]

We present a microscopic theory that describes the ordering of two distinct ligands on the surface of a faceted nanoparticle. The theory predicts that when one type of ligand is significantly bulkier than all others, the larger ligands preferentially align themselves along the edges and vertices of the nanoparticle. Monte Carlo simulations confirm these predictions. We show that the intrinsic conformational entropy of 10 the ligands stabilizes this novel edge-aligned phase.

## 1. Introduction

It is well known that nanoparticles (NPs) and block copolymers self-assemble into minimal surface, minimal energy structures. While these structures form quite reliably, they lack 15 the hierarchical complexity needed for the development of novel functional materials. The next generation of nanomaterials should not be limited to the simple symmetric patterns easily obtained by a monodisperse solution of identical particles. New structures should instead follow the lead of biology, whose primary design 20 goal is the ability to accomplish functional tasks. To accomplish its goals, biology produces molecules that are often asymmetric and contain various hierarchical levels of structure and complexity. Consider proteins. From the set of a mere 20 amino acids, all the molecular machinery necessary for living processes 25 can be assembled. Like proteins, the next generation of nanoscale building blocks should be required not merely to self-assemble, but to self-assemble into a diversity of structures with arbitrary shape and composition. Unfortunately, our present understanding is insufficient to determine how even a single protein folds much 30 less how it accomplishes complex tasks. While biology has the advantage of billions of year of evolutionary trial and error, researchers would like something more predictable and designable.

Inspired by recent advances in the synthesis of faceted 35 nanocrystals, [1-4] one potential route towards developing building blocks that contain preprogrammed instructions for switchable[5] homogeneous and heterogeneous[6] self-assembly is the patterning of selective, directional, attractive "patches" directly onto particle surfaces.[7-13] Several examples of functionalizing 40 particles in this way exist in the literature. [14-20] For example, as detailed in ref. 5, reconfigurable three-dimensional nanoparticle superlattices have been constructed using a DNA patch as a linker between particles. Imagine functionalizing a nanocube with multiple ligand patches. Now consider the assembly advantages 45 one would gain by isolating each ligand patch on a different face of the cube (Fig. 1A). Through judicious choice of ligands,

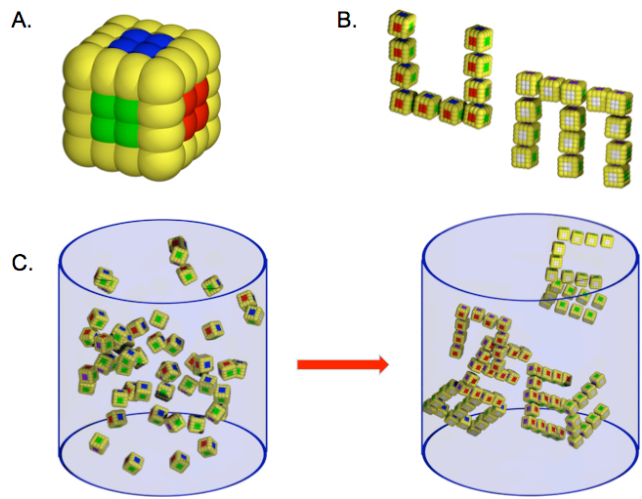

60 ligand binding interactions would restrict particles to preferentially bind to other particles by aligning faces compartmentalized with like or complementary ligands. With this specificity, one could, by appropriately choosing the location of the patches, program particles to assemble into structures of 65 arbitrary complexity (Fig. 1B-C). Such an advance would represent an important step towards designing structures tailor-made for accomplishing functional tasks. As a first step, we investigate in this paper the patterning of ligand-coated facetted nanoparticles to understand how one might achieve such ligand compartmentalization.

Self-assembled monolayers (SAMs) formed by the adsorption of functionalized molecular chains are intensely studied due to their unique surface chemical properties.[21-23] In addition to offering unique electronic and optical properties, SAM-coated nanoparticles (NPs) may be candidates for the type of novel programmable patchy building blocks described above.[8,13,24] Consequently, we seek to develop new ways to control the arrangement of molecular surfactants on NPs. In practice, it is of course difficult to dictate *a priori* the precise location at which a ligand will attach to a particle.[25] An important step towards synthesizing patchy particles with unusual ligand patterns [26-30] was realized several years ago, when a mixture of molecules that comprised a ligand shell on the surface of a gold nanoparticle was found to phase separate into ordered domains 5 Å in width.[31] More recently, simulations of ligands bound to the surface of a cylindrical, rod-shaped nanoparticle found that entropic effects could lead to the patterning of stripes around the circumference of the rods.[32,33,34] While these results represent novel and important advances, still more control is desired.

Entropic stabilization is one promising route towards controlling the position of patches on nanoparticles. There are many examples of entropy-driven ordering. Alder and Wainwright, and Wood and Jacobson showed that hard spheres crystallize into an FCC lattice because this phase has the highest entropy.[35-38] Liquid crystal and plastic crystal mesophases are possible in nonspherical hard particles.[39,40,41] Recently, Haji-Akbari *et al.* demonstrated that entropy orders hard tetrahedrons into quasicrystals.[42] The stable phases for a system of soft spheres are determined by a competition between a close-packing rule associated with positional entropy of the particles and a minimal-area rule associated with the internal entropy of the soft coronas.[43] Entropic contributions also play a large role in determining the stable phases of polymers confined to certain regions.[44,45] Recently, it was shown that a binary system of tethers bound to a planar surface could form stripes when the conformational entropy of the tethers in this phase dominates the configurational entropy that can be obtained in the mixed state.[46]

In this paper, we illustrate how entropic stabilization can be exploited to selectively order ligands on a polyhedrally-shaped NP's vertices, edges, and faces. Recently, DNA-linkers grafted to a nanocrystal have been observed to preferentially bind onto regions of higher surface curvature due to entropic contributions.[47] Using a simple coarse-grained Ising-like model, we conducted theoretical calculations and numerical simulations to predict the entropically-driven phase separation of two immiscible ligands into distinct patches on the vertex, edge, and face regions of a faceted particle. We confirmed these results by simulating, using Monte Carlo methods, a second model that explicitly includes ligand-ligand interactions. Section 2 presents the theory, Section 3 presents the simulation results and discusses them in the context of the theoretical predictions, and we conclude in Section 4.

## 2. Theory

The synthesis of faceted nanoparticles with sizes ranging from 10-20 nm has been well demonstrated in the literature.[48,49] In addition, we posit the existence of some chemically attachable ligand. If we assume nearly complete surface coverage by a 5 Å wide ligand molecule, we obtain on the order of 10 strands per particle edge or roughly $10^2$-$10^3$ total ligands bound to the particle's surface. We consider two such ligands A and B, where ligand A is both chemically distinct from, and sterically bulkier than, ligand B, as illustrated in Fig. 2A. The extra bulk of ligand A may arise by choosing it to be a branched molecule. We assume that ligand A binds to the particle with the same affinity as B. The ratio of the two ligands adsorbed onto the particle can be controlled by varying the stoichiometry of the reagents during the synthesis.[30] A complete description of the adsorption process is beyond the scope of this paper, but it should be noted that in the absence of cooperative effects, the ratio of tethers adsorbed for a reasonably large particle should be approximately determined by the central limit theorem. For example, if tether B has the same binding strength as tether A but composes only 30% of the ligands in solution, then of 500 bound tethers approximately 30% or 150 ligands should be of type B. Fluctuations should scale as the square root of the mean leading to a standard deviation of about 12 tethers or 8% of the mean. These fluctuations should decrease as the number of tethers on the surface increases.

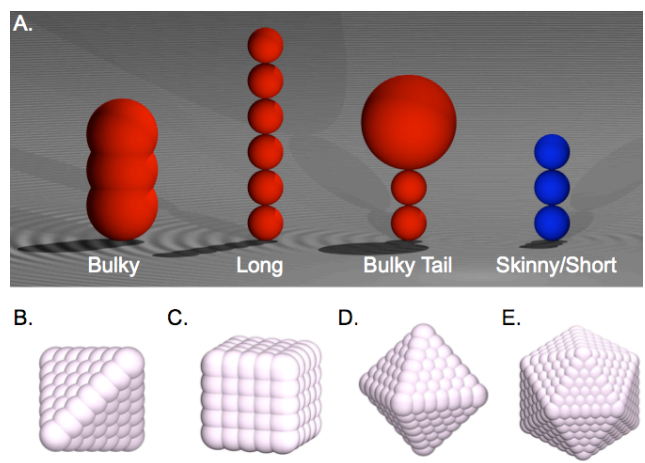

**Fig. 2** (A) Various tether types used in simulations. Throughout the paper larger more sterically hindered tethers are shown in red and labeled as type A, while smaller less sterically hindered tethers are shown in blue an referred to as type B. Lattice binding sites for a (B) tetrahedron, (C) cube, (D) octahedron, and (E) icosahedron. Because of their inherent symmetries, tetrahedrons, octahedrons, and icosahedrons are meshed by a triangular lattice and cubes are meshed by a square lattice.

We consider a highly coarse-grained model for the system described above. We first construct a lattice of binding sites on the surfaces of a tetrahedron, cube, octahedron, and icosahedron. Given the symmetry present in each face, we use triangular lattices for the tetrahedron, octahedron, and icosahedron, and a square lattice for the cube as shown in Fig. 2B-E. Each lattice site is represented by a bead of radius $\sigma_L = 1$. Nearest-neighbor sites are separated by a distance $\sigma_L$. Each site has tethered to it one ligand of either type A or type B. Depending on the choice of ligand, there may be a net attraction between like species due to van der Waals or hydrophilic/hydrophobic forces. The Hamiltonian for this system can be written as,

$$H = \sum_{\langle ij \rangle} E_{ij}, \qquad (1)$$

where the sum is over nearest-neighbors and the indices $i$ and $j$ denote the different species of tether beads on the tails of the ligands. Identical nearest-neighbor ligands interact with energy $E_{AA} = E_{BB} = \xi$ while unlike neighbors interact with energy $E_{AB} = 0$. Here, $\xi < 0$ provides an attraction between like ligands.

Under the criteria listed above, our model is similar to an Ising model in that each site can exist in one of two states that are self-attracting. However, the major distinction between this and a standard Ising model is that here individual "spins" contain an intrinsic degeneracy because there are multiple conformations accessible to the tether that the spin variable represents. The thermodynamically preferred macrostate will be determined by a competition between the internal energy, the intrinsic conformational entropy, and the configurational entropy arising from the arrangement of ligand tethers on the surface.

Consider an arbitrary convex faceted particle with $N$ binding sites. There are $N_V$, $N_E$, and $N_F$ binding sites located on the particle's vertices, edges, and faces, respectively, such that

$$N_V + N_E + N_F = N. \qquad (2)$$

Formulas for the values of $N$, $N_V$, $N_E$, and $N_F$ for a convex polyhedron of a given edge length $L_E$ are given in Table 1 for each of the regular polyhedrons considered. We restrict the number ratio of the two ligands to remain fixed and define the fraction of type A ligands to be $\phi_A$ and the fraction of type B ligands to be $\phi_B = 1 - \phi_A$. We further define $N_A = \phi_A N$ and $N_B = N - N_A$ as the number of ligands of type A and B, respectively.

Recent experimental studies have shown that ligands and small particles preferentially bind to edges and vertices.[50,51] We are interested in states where bulky ligand tethers preferentially distribute themselves along the edges and vertices of the nanoparticle. For this reason, we divide this system into a vertex-edge region and a face region. We define $N_{A,VE}$ ($N_{A,F}$) as the number of type A ligands found on the vertices and edges (faces). For particles with well-defined binding sites, the configurational entropy of a system with a known value of $N_{A,VE}$ can be computed exactly by counting the number of microstates. The number of unique ways one can arrange $N_{A,VE}$ ligands of type A onto $N_{VE} = N_V + N_E$ sites is given by,

$$\Omega_{VE} = \frac{N_{VE}!}{N_{A,VE}!\left(N_{VE} - N_{A,VE}\right)}. \qquad (3)$$

Likewise, the number of ways of arranging $N_{A,F}$ ligands of type A onto $N_F$ sites is given by,

$$\Omega_F = \frac{N_F!}{N_{A,F}!\left(N_F - N_{A,F}\right)} = \frac{N_F!}{\left(N_A - N_{A,VE}\right)\left(N_F - N_A + N_{A,VE}\right)}. \qquad (4)$$

The configurational entropy for a system with $N_{A,VE}$ ligands of type A bonded along the edges and vertices is then,

$$S_{config}\left(N_{A,VE}\right) = k_B \ln\left[\Omega_{VE} \cdot \Omega_F\right], \qquad (5)$$

where $k_B$ is Boltzmann's constant. For a macrostate that is defined for multiple values of $N_{A,VE}$, the total entropy can be computed by summing $S_{config}\left(N_{A,VE}\right)$ over all $N_{A,VE}$ consistent with the macrostate.

The conformational entropy of a ligand tether will depend on its length, degree of branching, the radius of the beads used to model the molecular groups in the tether, the configuration of ligands on the surface, and the type of site to which a ligand is adsorbed (i.e. vertex, edge, or face.) Vertex sites will have more free volume than edge sites, which will have more free volume than face sites. This results in more possible conformations obtainable by tethers on vertex and edge sites. We define the number of conformations accessible to an isolated ligand of type $i$ bound to a site of type $x$ to be $\Omega_{x,i}$. In general, $\Omega_{x,i}$ will not be finite since a ligand tether's beads are not confined to a lattice. This is of little importance however, since only the ratio of conformations accessible to different sites is needed to compute differences in the entropy. In addition to the type of site, the number of accessible conformations available to a tether is restricted by the configuration of other tethers in its local vicinity. In general, two closely spaced tethers will reduce each other's free volume, thereby restricting the number of accessible conformations. The interactions between a set of tethers may be viewed as a complicated many-body problem.[46] We define the avoidance probability $p(i \otimes \{k\})$ as the probability that tether $i$ will not intersect any tethers in the set $\{k\}$ found within its local vicinity. We define the local vicinity to be the maximum range over which tethers can overlap. Here, $\{k\}$ is a set of self- and mutually-avoiding tethers. From the avoidance probability, one can obtain the conformational entropy of tether $i$,

$$S_{conform,i} = k_B \ln\left[\Omega_{x,i} p(i \otimes \{k\})\right], \qquad (6)$$

where the total conformational entropy of the tethers on the particle is given by,

$$S_{conform} = \sum_{i=0}^{N-1} S_{conform,i}. \qquad (7)$$

The free energy of a system with a given configuration is then given by,

$$F = H - TS_{conform} \qquad (8)$$

To compute the change in entropy that occurs when tether $i$ moves from site $y$ with a local configuration of tethers $\{j\}$ to another site $x$ with a local configuration of tethers $\{k\}$, we have

$$\frac{\Delta S_i}{k_B} = \ln\left[\Omega_{x,i} p(i \otimes \{k\})\right] - \ln\left[\Omega_{y,i} p(i \otimes \{j\})\right]$$
$$= \ln\left[\frac{\Omega_{x,i} p(i \otimes \{k\})}{\Omega_{y,i} p(i \otimes \{j\})}\right]. \qquad (9)$$

In a previous publication,[33] we demonstrated that one can quickly obtain qualitatively correct phase diagrams for a binary system of

# ARTICLE TYPE

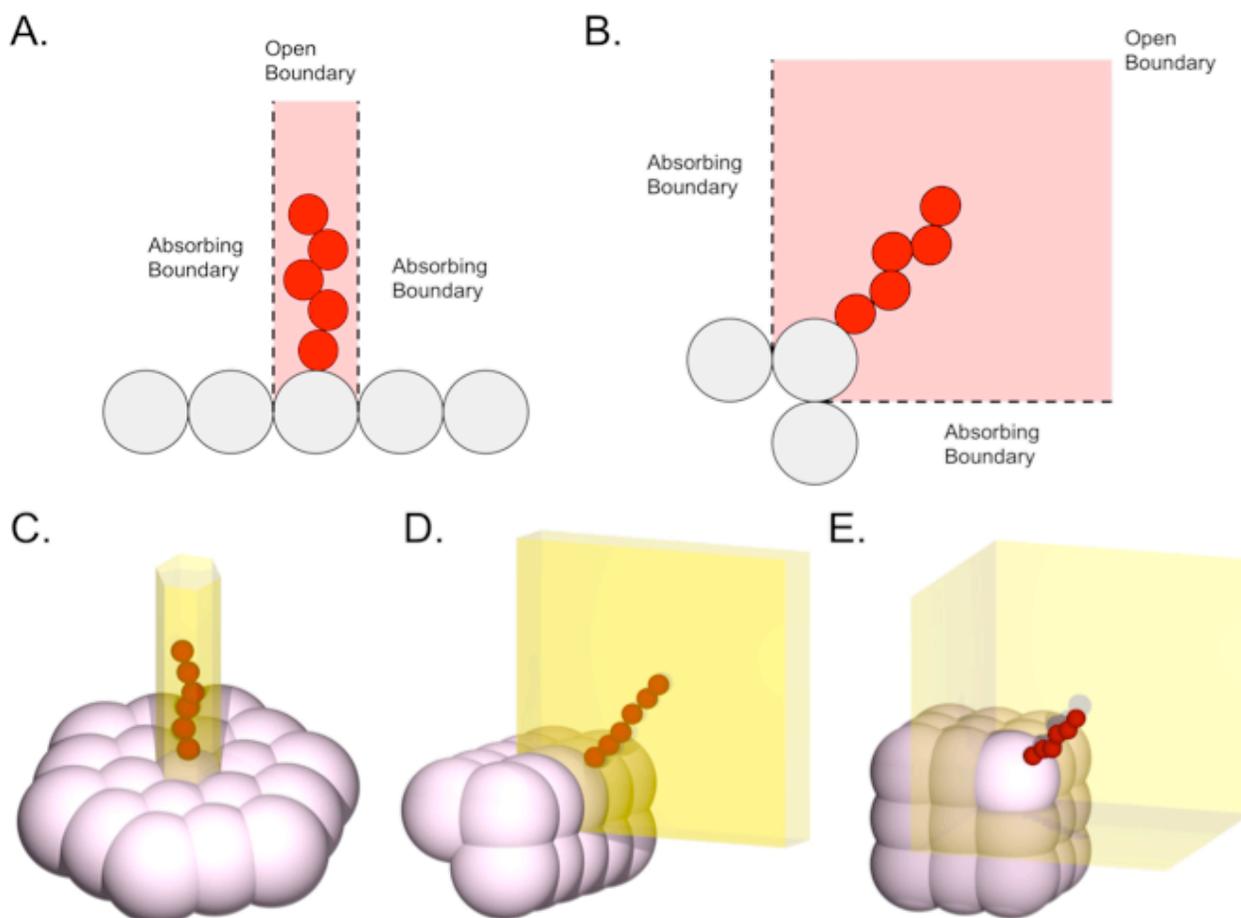

**Fig.3** Voronoi cells for different types of sites. (A) Here, we show schematically the two-dimensional analog of the Voronoi cell for a face site. The cell is bound on either side by two parallel walls, but is unbound above. (B) In contrast, the Voronoi cell for an edge site has perpendicular absorbing boundaries providing more free volume than is available on a face site. (C) In three dimensions, the Voronoi cell defined for a face site (shown here for a
5 triangular lattice) sharply restricts tether conformations to remain within a narrow tube. This will not be the case for faceted particles. (D) The Voronoi cell defined for an edge site (shown here for a cube) allows tethers to sample the space in a wedge between two parallel planes. As can be seen, this confinement is less stringent than for faces. (E) In contrast to edge and vertex sites, the Voronoi cell defined for a vertex site (shown here for a cube) allows tethers to sample conformations in a much larger volume.

10 ligands tethered to a planar surface by approximating the many-body tether interaction as a two body interaction between tethers. This two-body approximation relies heavily on the fact that the surface is isotropic, since this allows the potential between tethers to be determined by their separation alone. This will not be the
15 case for faceted particles. In principle, one can still compute the ratio of accessible conformations using a Monte Carlo integration scheme, but the interaction between tethers will now depend on their location and orientation relative to the edges of the particle. The complexity of the new potential makes finding a more
20 suitable approximation for the conformational entropy desirable.

To simplify the problem, we used a simple free volume approximation to compute the conformational entropy of a tether located on any of the three types of sites. In this approximation, we replaced the interactions between particles with an effective

25 field that interacts differently with tethers at different sites. To zeroth order, one can approximate the free volume accessible to a tether as being equivalent to the volume of the Voronoi cell constructed around its lattice binding site as illustrated in Fig. 3. Because the particle is convex, the space defined by the Voronoi
30 cell will not be bounded above. The Voronoi cell acts as an effective field by restricting a tether's movement. This field is defined so that all tether conformations that remain within the Voronoi cell are allowed while those that penetrate the Voronoi cell are not allowed. Under this assumption, the survival
35 probability is equivalent to the fraction $f_{x,i}$ of conformations accessible to a tether of type $i$ located at site $x$,

$$p(i \otimes \{k\}) \approx f_{x,i}. \qquad (10)$$

Solving for the fraction of accessible tether conformations is then equivalent to finding the survival probability for self-avoiding random walks (SAWs) starting at the origin that remain within a Weyl chamber defined by the Voronoi cell.[52]

Random walks (RWs) confined to a wedge have been considered previously in the literature. [52-56] To the best of our knowledge, no one has yet considered three dimensional SAWs confined to a Weyl chamber. SAWs are non-Markovian, and although much is known about them, solving them is still beyond the capability of current combinatoric techniques.[52] In principle, it may be possible to obtain an analytic form for Markovian RWs confined to Weyl chambers where the absorbing walls are not orthogonal to each other, but this problem is far from trivial. One could easily solve a master equation for these boundary conditions computationally, but if a computational approach is to be undertaken it is best to include self-avoidance since this more accurately reflects the physics of a tether. For this reason, we used Monte Carlo integration to numerically compute the fraction $f_{x,i}$ of tethers of type $i$ bonded at site $x$ that remain inside that site's Voronoi cell. This calculation was done off-lattice in the following way. First, we constructed Voronoi cells for each type of lattice site for each of the polyhedron lattices described above. Next, we placed the first tether bead at a random point on a sphere of radius $\sigma_L + d/2$ centered on the binding site. Here, $d = 0.5$ is the separation between adjacent tether beads. Additional tether beads were added to the sequence by placing them at a random point on a sphere of radius $d$ centered on the previous bead. Each tether contained $L = 3$ beads. All tether conformations in which beads overlap (i.e. the separation between beads was less than $d$) were rejected since these are not self-avoiding. We used the above procedure to generate $10^9$ unique self-avoiding tether conformations. Using the bead positions from these conformations, we then varied the bead radius $\sigma_{T,i}$ and computed the fraction of tether conformations that remain inside the Voronoi cell as a function of $\sigma_{T,i}$. A plot of the conformational entropy vs. bead radius is given in Fig. 4 for each type of site found on each of the lattices. As should be expected, the vertex sites for each lattice always have the largest fraction of accessible conformations followed by edge sites and then face sites. The difference in entropy between sites grows as one increases the bead radius. Since the fraction of conformations accessible to tethers at edges and faces asymptotically approaches zero as $\sigma_{T,i} \to 0.5$, the entropy will diverge to minus infinity at this point. This indicates that one can greatly increase the entropy difference between different sites on a lattice by increasing the tether bead radius.

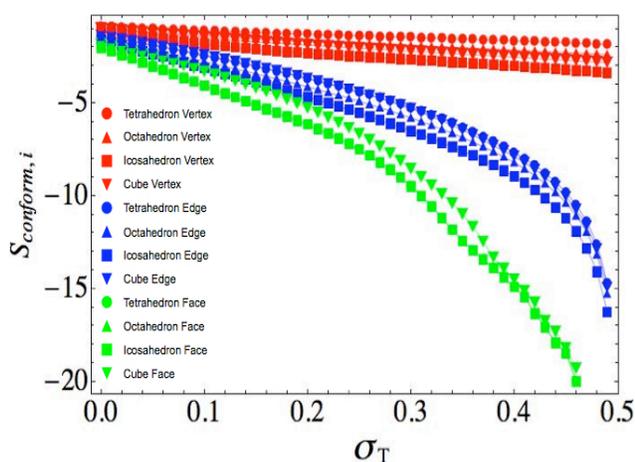

**Fig. 4** A plot of the conformational entropy $S_{conform,i}$ of a tether in units of $k_B$ obtained via MC integration of a SAW as a function bead radius for vertex (red), edge (blue), and face (green) sites on tetrahedrons (disks), octahedrons (triangle), icosahedrons (square), and cubes (inverted triangles). All tethers used are 3 beads long.

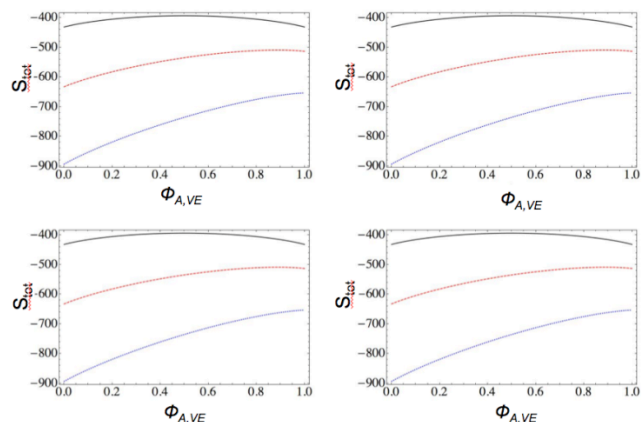

**Fig. 5** Plots of the total entropy $S_{tot}$ as a function of $\Phi_{A,VE}$ for the (A) tetrahedron, (B) cube, (C) octahedron, and (D) icosahedron. For each plot, $\Phi_A = 0.5$ so that there are an equal number of type A and type B tethers. In each case, we set the bead radius of type B tethers equal 0.25. The type A tether bead radius was chosen to be 0.25 (black, solid), 0.36 (red, dashed), and 0.47 (blue, dotted). As the radius of type A beads increases, the maximum entropy configuration shifts from $\Phi_{A,VE} = 0.5$ to $\Phi_{A,VE} = 1.0$.

The stable configuration of the system can be determined by minimizing the free energy. In the absence of energetic interactions ($\xi = 0$), the stable configuration will maximize the total entropy

$$S_{tot} = S_{config} + S_{conform} \qquad (11)$$

In Fig. 5, we plot the total entropy $S_{tot}$ as a function of $\phi_{A,VE} = N_{A,VE}/N_{VE}$ for each of the particle geometries shown in Fig. 2B-E. For each system, we chose $\phi_A = 0.5$. In each case, the entropy is maximized at $\phi_{A,VE} = \phi_A$ when tethers A and B have the same bead radius, suggesting no preference for alignment along the edges. As the radius of type A beads increases, the maximum entropy shifts to configurations with large fractions of type A tethers on the edges indicating that entropic effects cause bulky tethers to align along the edges and vertices of the particle. This edge-aligned state, if it can be achieved experimentally, is of great interest because it provides a simple way to controllably position patches on a nanoparticle.

The analysis of the simple model presented above only applies when the immiscibility between tethers is negligible. In real systems, the immiscibility depends on the types of ligands and is often quite large. Moreover, it is not clear whether the

# ARTICLE TYPE

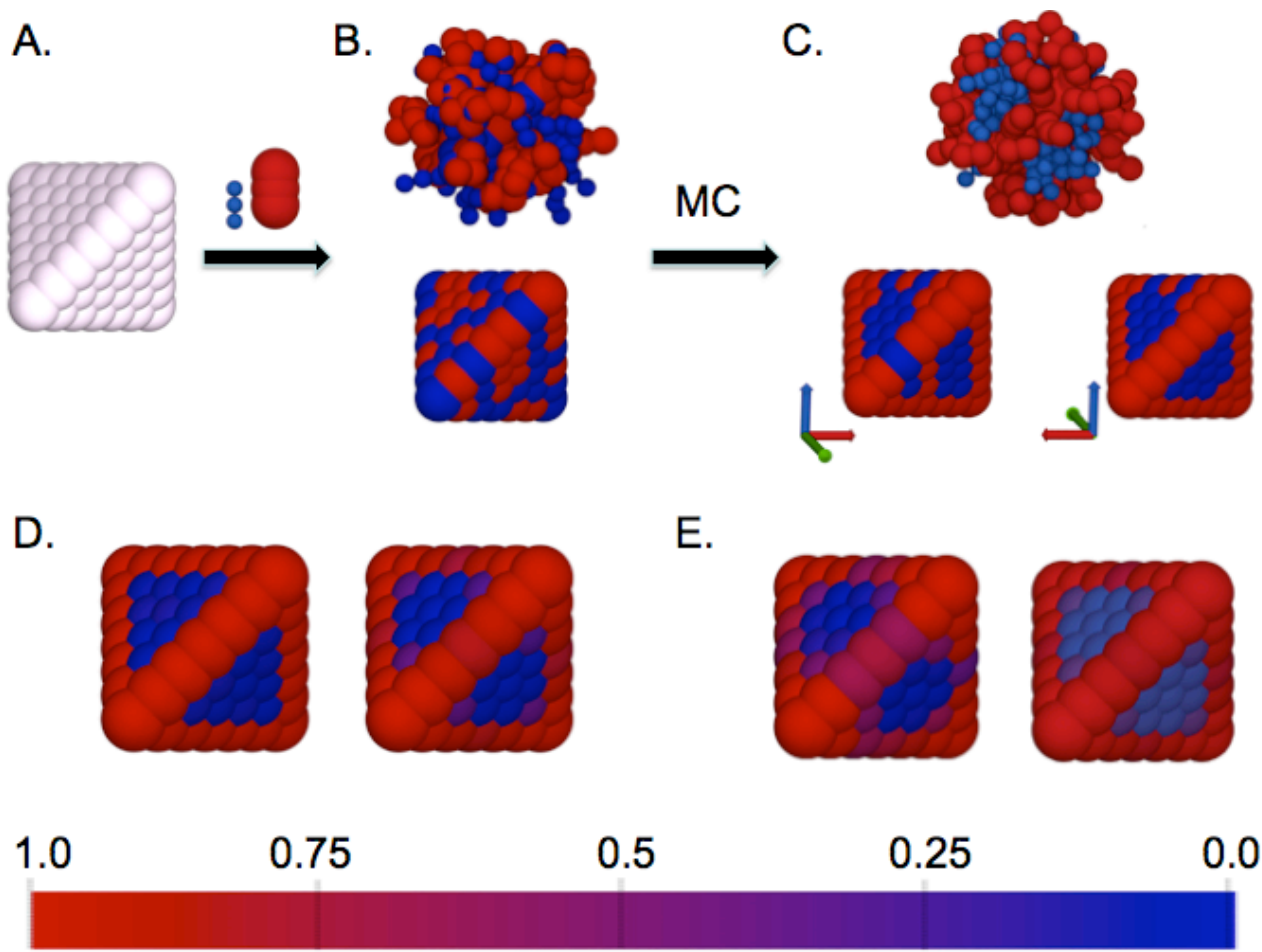

**Fig.6.** (A) All simulations are initialized by randomly arranging bulky (red) and skinny (blue) tethers on the lattice. (B) A single starting configuration of the explicit-tether model is shown both with the tethers included (top) and with the tethers removed for easier viewing (bottom). (C) When run with the Monte Carlo algorithm, bulky tethers will preferentially assemble along the edges and vertices. Again, we show a single snapshot of the system both with
5 tethers (top) and without tethers for better viewing (bottom). Both the front (bottom-left) and back (bottom-right) views are shown. (D) To test the effect of energetic interactions, we ran the implicit-tether models both with (right) and without (left) immiscibility for $\sigma_{T,A} = 0.46$ and $\sigma_{T,B} = 0.25$ and same number of beads $L_i = 3$. These images represent ensemble averages. The color scale on the bottom measures the fraction of type A tethers found on the sites with 1.0 corresponding to all type A tethers and 0.0 corresponding to all type B tethers. Both miscible and immiscible cases exhibit preferential alignment of the bulky tethers along the edges and vertices. In the presence of immiscibility, there is a slight preference for bulky tethers to appear at the face site
10 nearest to a vertex. This is to be expected since this site typically features three type A neighbors, which makes it an energetically favorable site. (E) Similar results are obtained with the explicit-tether model for systems with immiscible (right) and miscible (left) interactions for $\sigma_{T,A} = 1.5$ , $\sigma_{T,B} = 0.5$ and same number of beads for both tehters $L_i = 3$ . In this model, the immiscible system gives slightly better alignment of the bulky tethers along the edges and vertices.

Voronoi cell approximation is valid since it neglects the effect
15 that the local configuration will have on a tether's conformational entropy. Below, we use Monte Carlo simulations to validate the Voronoi cell approximation and show that even in the presence of immiscibility and local configuration effects, the edge-aligned state is still thermodynamically preferred.

## 3. Simulations and Discussion

20 To determine how immiscibility and the local configuration affect the results of the theory presented above, we simulated the self-assembly of tethers on faceted surfaces using both the implicit-tether model described in the theory section and an explicit-tether
25 model to be described shortly. Simulations of both models are initialized with the tethers arranged randomly on the lattice surface and then run using Metropolis Monte Carlo (MC) algorithm with Kawasaki dynamics (Fig. 6A-C). In this algorithm, trial moves are generated by choosing two lattice sites
30 at random and swapping their tethers. The change in the energy $\Delta E$ is then computed. Since the implicit-tether model includes a conformation entropy term, the change in entropy $\Delta S$ must be computed as well. Trial moves are accepted with a probability proportional to min[1, exp{−β(ΔE − ΔS)}], where β is the inverse temperature. The inclusion of the entropy in Boltzmann weight arises when there is degeneracy in states of the system.[46] This acceptance probability ensures that detailed balance holds so that the system will sample an equilibrium distribution of configurations.

Since the tetrahedron has the smallest number of sites and is therefore fastest to simulate, we chose it as the canonical example on which to test the self-assembly of tethers. Using the lattice shown in Fig. 2B, we simulated the implicit-tether model both with and without immiscibility. In this model, information about the tethers' conformations has been coarse-grained out and replaced with a conformational entropy term. The configuration of tethers on the surface completely describes the state of the system. In this way, the system is similar to an Ising model in that the state can be completely specified by a set of "spins" where up and down spins correspond to type A and type B tethers. Unlike true Ising spins, each tether exhibits intrinsic conformational entropy that depends on the type of tether and the type of site. The conformational entropy of tethers A and B corresponds to tethers with length $L_i = 3$ and bead radius $\sigma_{T,i} = 0.46$ and 0.25, respectively. We chose $\beta\xi = 0$ for miscible tethers and $\beta\xi = −1$ for immiscible tethers. Changes in the energy and entropy were computed using equations (1) and (9), respectively. Each system was initialized with $\phi_A = 0.5$ and tethers arranged randomly on the surface. We ran all systems for 20,000 MC steps until the energy had equilibrated. In Fig. 6D, we show the ensemble average results of 100 runs for both the miscible and immiscible cases. From the figure it is clear that the system can tolerate a significant amount of immiscibility and still remain in an edge-aligned state.

While the simulations above suggest that edge-aligned states are stable even in the presence of a modest amount of immiscibility between unlike tethers, they still rely on the approximate Voronoi cell free volume calculations of the conformational entropy. These calculations assumed that a tether's conformational entropy depends much more strongly on the type of lattice site (i.e. vertex, edge, or face) than it does on the local configuration of tethers. To check that this assumption holds, we ran MC simulations for this system using a model that explicitly contains the tethers' conformations. In this explicit-tether model, a tether is again represented by a set of spherical beads of radius $\sigma_{T,i}$ and conformations are generated in the same manner described above. Beads can be bonded in a variety of ways to represent different ligand morphologies. To model immiscibility, beads from different tethers interact through a soft potential of the form

$$V(r_{ij}) = \begin{cases} V_{ij} \cdot \dfrac{(R_{ij} - r_{ij})}{R_{ij}} & \text{if } R_{ij} \geq r_{ij} \\ 0 & \text{if } R_{ij} < r_{ij} \end{cases} \quad (12)$$

where $r_{ij}$ is the separation between beads, $V_{ij}$ is the strength of the potential, and $R_{ij} = \sigma_{T,i} + \sigma_{T,j}$ is the range of interaction. The use of a soft potential is advantageous here because it increases the acceptance rate of trial moves. This potential has previously been used in dissipative particle dynamic simulations of ligands tethered to the surface of spheres, cylinders, and planes and has successfully predicted the formation of stripes on these surfaces.[31,32,33] The value of $\beta V_{ij}$ was chosen to be 25 for like tethers and 40 for dissimilar tethers, consistent with previous simulations.[32,33] We generated trial moves by randomly choosing two lattice sites and swapping their tethers. During the swap, new self-avoiding conformations are chosen for each tether. Moves were then accepted or rejected with the MC criteria. Since the conformations are known explicitly in this model, there is no degeneracy in any state, and $\Delta S = 0$ for all trial moves.

In order to compare the explicit- and implicit-tether models, we first simulated miscible and immiscible explicit-tether model with bulky ($L_i = 3$, $\sigma_{T,i} = 1.5$) and skinny ($L_i = 3$, $\sigma_{T,i} = 0.5$) tethers. Since the potential is soft, we chose a larger radius than was used in the implicit-tether model, so that tether beads will overlap and influence each other. As before, systems were initialized randomly with $\phi_A = 0.5$. They were then equilibrated for $\geq 60,000$ MC steps after which the final state was recorded. The final configurations of $\geq 50$ runs were averaged and the results are shown in Fig. 6E. As can be seen, there is substantial migration of the larger type A tethers to the edges and vertices. This is largely consistent with predictions from the theoretical implicit-tether model. One small discrepancy occurs at the face site closest to the vertex for miscible systems. As can be seen from Figs. 6D and 6E, this bead is less likely to be occupied by bulky tethers in the implicit-tether model. This behavior is expected because the implicit-tether model approximates all face sites as having the same free volume. Under this approximation, all face sites have the same likelihood of binding to a bulky tether when the system is miscible. In a real system, face sites that are near an edge will have slightly more free volume than sites in the middle of a face, making bulky tethers more likely to bind there.

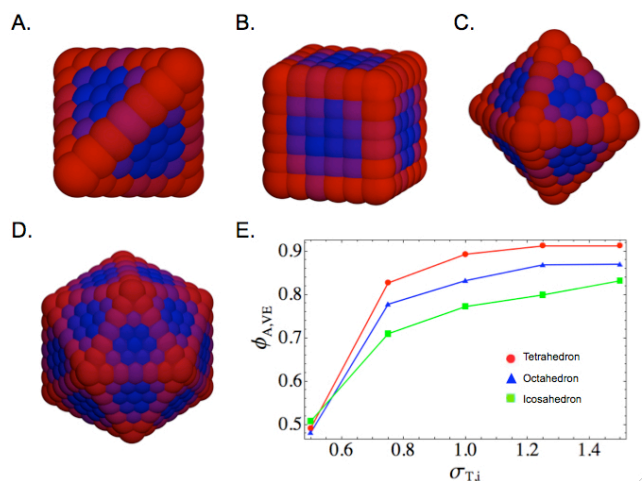

**Fig. 7** This figure represents an ensemble average of $\geq 50$ runs of the explicit-tether model with bulky-skinny tethers for the (A) tetrahedron, (B) cube, (C) octahedron, and (D) icosahedron. Tethers have been removed for easy viewing. (E) A plot of the fraction $\Phi_{A,VE}$ of vertex/edge sites with bulky tethers vs. the bead radius of the bulky tether $\sigma_{T,i}$ for the tetrahedron (red, circles), octahedron (blue, triangles), and icosahedron (green, square). As the sharpness of the edge increases (tetrahedron > octahedron > icosahedron), so does the likelihood that bulkier tethers reside there. (The cube is not shown in Figure E because its underlying square lattice is different than the other three geometries making direct comparisons impossible.)

The simulations described above indicate that neither modest amounts of immiscibility nor the local configuration of tethers is enough to disrupt the stability of the edge-aligned state for tetrahedrons. The edge-aligned-state is expected to be more stable on the tetrahedron than other regular polyhedron surfaces because its edges are the sharpest and provide the most additional free volume. To test whether or not edge-aligned states remain stable in other polyhedrons, we simulated bulky ($L_i = 3$, $\sigma_{T,i} = 1.5$) and skinny ($L_i = 3$, $\sigma_{T,j} = 0.5$) tethers for the tetrahedron, octahedron, cube, and icosahedron lattices using the explicit-tether model. As before, all systems were run with $\phi_A = 0.5$ from an initially random configuration, and systems were run until equilibrium was reached. The final configurations of $\geq 75$ runs were then averaged and the results are shown in Fig. 7. As can be seen, all geometries show significant migration of the bulky tethers to the edge-vertex region resulting in a mostly edge-aligned state. While there is clearly a preference for bulky tethers to align along the edges, it is worth noting that this effect is significantly weaker in the icosahedron. As one can see from Fig. 7D, the concentration of bulky tethers along the icosahedron edges is not nearly as large as the tetrahedron case. This is to be expected since adjoining icosahedron faces are at a more obtuse angle and consequently their common edge provides less free volume for the bulky tethers.

The theoretical prediction of edge-aligned states relies on a free volume argument. If this argument remains valid for systems that include both immiscibility and explicit information about tether conformations, one should observe two effects: (1) edge-aligned states will be stronger for bulkier tethers and (2) edge-alignment must get progressively weaker as the sharpness of the edges decreases. To test whether or not these statements hold, we ran the explicit tether model with skinny ($L_i = 3$, $\sigma_{T,j} = 0.5$) tethers and tethers of variable radius ($L_i = 3$, $\sigma_{T,j} = 0.5 - 1.5$) for the tetrahedron, octahedron, and icosahedron lattices. Cubes were not considered because the underlying square lattice has a different symmetry than the other three shapes and would not be an appropriate comparison. We ran each system $\geq 20$ times and computed the average fraction $\phi_{A,VE}$ of bulky tethers on edges and vertices. A plot of $\phi_{A,VE}$ vs. the bulky tether's bead radius is given in Fig. 7E. As can be seen, increasing the radius of the bulky tether results in a greater fraction of bulky tethers on the edges and vertices. In addition, $\phi_{A,VE}$ decreases as the shape changes from tetrahedron to octahedron to icosahedron, indicating that sharper edges indeed enhance edge-alignment. These results are in accordance with the free volume argument above and suggest that edge-alignment is due to entropic effects that arise from restricting the free volume accessible to tethers.

Experimental systems are unlikely to have an exactly 50/50 ratio of bulky-to-skinny ligands adsorbed on the surface. For this reason, we simulated systems whose concentration of ligands is slightly asymmetric to see how the ratio of components affects the morphology of the equilibrium structures. As before, we used the explicit-tether model with bulky ($L_i = 3$, $\sigma_{T,j} = 1.5$) and skinny ($L_i = 3$, $\sigma_{T,j} = 0.5$) tethers. In Fig. 8, we show the ensemble averages obtained from 100 MC runs of systems with asymmetric compositions in the range $\phi_A = 0.46 - 0.54$. We find that the fraction $\phi_{A,VE}$ of vertex-edge sites occupied by a bulky tether monotonically increases as $\phi_A$ increases. This is not noticeable at the midsection of the edges, suggesting that the loss entirely unexpected, since the overall concentration of bulky tethers is increasing. The effect is, however, especially of free volume strengthens entropic effects and enhances the migration of bulky tethers towards the edges. In this interpretation, additional bulky tethers on face sites reduce the conformational entropy of their neighbours, especially if other bulky tethers are in the vicinity. As a result, bulky tethers are forced to occupy the edges and vertices where they can both retain more conformational entropy and reduce the repulsive interactions between immiscible species.

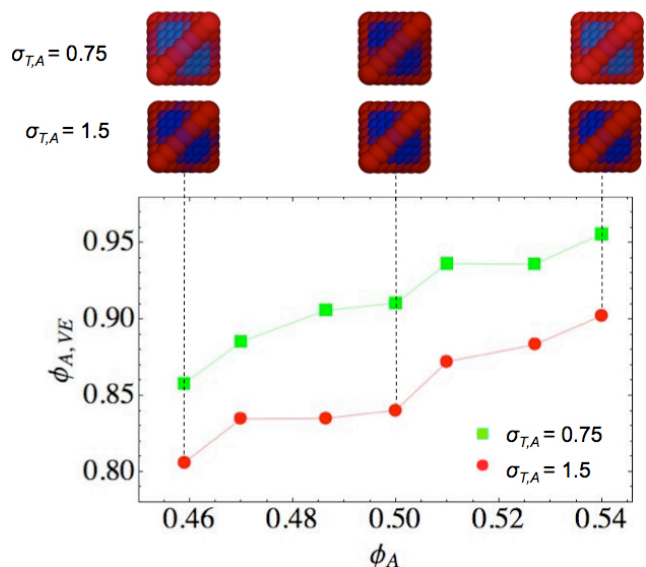

**Fig.8** A plot of the fraction of edge/vertex sites containing a bulky tether vs. the fraction of type A tethers on the tetrahedron. Data are shown $\sigma_{T,A} = 0.75$ and $\sigma_{T,B} = 0.5$ (green squares), and for $\sigma_{T,A} = 1.5$ and $\sigma_{T,B} = 0.5$ (red circles). Data points were taken from the average of 75 runs. The plot is accompanied by a picture of the ensemble averages for $\sigma_{T,A} = 0.75$ (top row) and $\sigma_{T,A} = 1.5$ (bottom row) for bulky tethers fractions $\phi_A = 0.46$, 0.5, and 0.54.

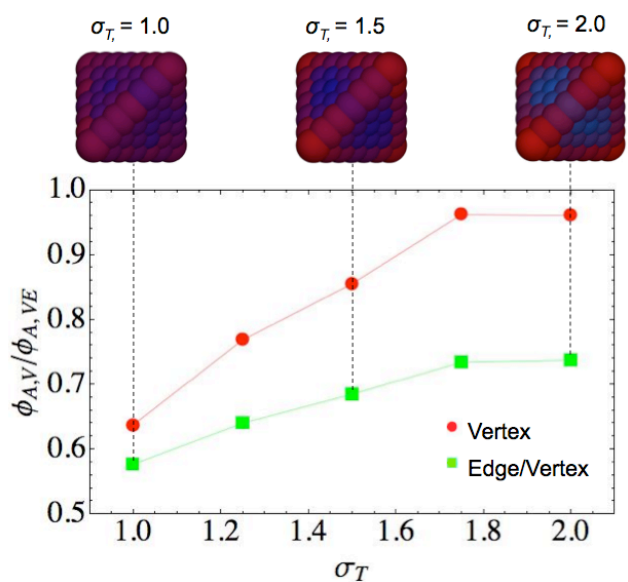

**Fig. 9** A plot of the fraction of bulky-tailed tethers found on edge/vertex (green squares) and vertex sites (red circles) vs. the end bead radius for the tetrahedron lattice, accompanied with ensemble average (upper snapshots) for different bulky-tailed tethers as indicated. Data points were averaged from 75 runs.

To examine the effect that a tether's shape has on the surface morphology, we simulated the explicit-tether model using the "bulky-tail/skinny" and "long/short" ligand tether geometries illustrated in Fig. 2A. In previous studies, these tether geometries self-assemble into striped patterns when adsorbed on flat, spherical, or cylindrical surfaces.[32,33,46] In our simulations, the long and short tethers contain $L_i = 6$ and $L_i = 3$ beads, respectively, each with a radius $\sigma_{T,i} = 0.75$. For the bulky-tail system, the tethers had identical length ($L_i = 3$) and radius ($\sigma_{T,i} = 0.75$), except for the bulky tether's tail bead, whose radius

$\sigma_T$ was varied from 1.0 to 2.0. All systems were first simulated on the tetrahedron NP, started from an initially random configuration with $\phi_A = 0.5$, and equilibrated for over 3000 MC steps. The long/short system exhibited a slight migration of long tethers towards vertices, but the effect is very weak compared to the bulky/skinny tether system. In contrast, the bulky-tail system showed some migration of the bulky tail tethers towards both the edges and vertices, but the effect was much weaker for the edges (Fig. 9). In both systems, the edge alignment is weaker than it is for systems with uniformly bulky tethers. These results indicate that stronger edge alignment occurs when bulkiness is introduced along the whole length of one of the ligand species. This effect should be expected since there is less total volume close to the particle surface. Near the surface, tether beads are more "squeezed" and free volume effects will be enhanced. By increasing the tether's bulkiness away from the particle surface, this effect is diminished because the bulky groups still have sufficient free volume and their conformational entropy does not decrease appreciably when they instead occupy a face site. As mentioned in the introduction, the synthesis of faceted patchy particles with multiple, distinct ligand patches would represent an important step towards the fabrication of programmable particles that could self-assemble into arbitrary shapes. To examine whether or not whether edge-alignment could be exploited to compartmentalize multiple ligand species, we simulated the explicit-tether model with a third tether on a larger tetrahedron NP comprised of 100 lattice beads. This system contains one bulky ( $L_i = 3$ , $\sigma_{T,i} = 1.95$ ) and two skinny ( $L_i = 3$ , $\sigma_{T,j} = 1.1$ ) tethers, each of which is immiscible with the others. The value of $\beta V_{ij}$ was chosen to be 35 for like tethers and 45 for dissimilar tethers. The two skinny tethers represent two different chemical ligand patches. The bulky tether's sole purpose is to "corral", and ultimately compartmentalize, the two skinny ligand species into patches on the faces. For this system, we chose the concentration of bulky tethers to be $\phi_A = 0.4$ , with equal amounts of the two skinny tether species. The system is initialized with the tethers arranged randomly on the surface and then run for 600,000 MC steps. From Fig. 10, we see that patches of a single species form on each face. This suggests that immiscible ligands could be compartmentalized to the face of a nanoparticle by adding a third bulky ligand to the mix.

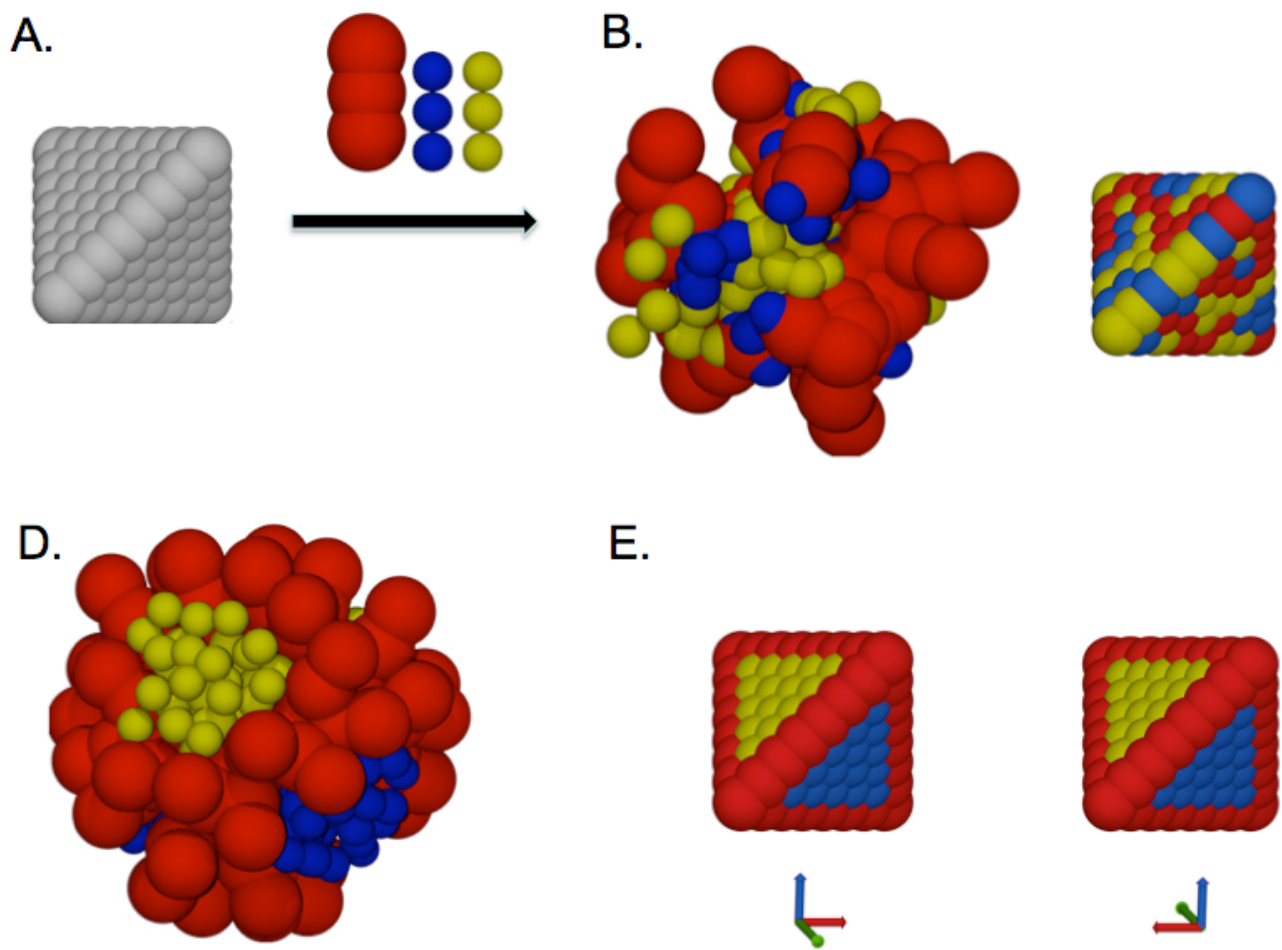

**Fig.10** (A) Ternary immiscible systems were simulated by adding a second skinny immiscible tether (yellow beads) to the tetrahedron system in the explicit-tether model. The bead radius for the bulky and skinny tethers are 1.5 and 1.1, respectively, and same number of beads $L_i$ =3. (B) An initial configuration of the model is shown both with tethers (left) and with the tethers removed for easy viewing (right). (C) In MC simulations, the bulky tethers preferentially migrate towards the edges and vertices, while the skinny tethers migrate towards the faces. can be seen, ligand patches are corralled on the faces and fenced in by a line of bulky tethers. (E) For easy viewing, we removed the tethers and show both the front (left) and back (right) views.

As stated earlier, our coarse-grained theoretical model is essentially the Ising model with an additional entropic term and boundary conditions that have been deformed to match that of a faceted surface. However, unlike the Ising model, which forms disordered and demixed phases, our coarse-grained model exhibits an ordered (here, edge-aligned) phase because of the additional conformational entropy available to tethers. Because the difference in conformational entropy between different types of sites increases sharply as the radius of the tether beads increases (Fig. 4), bulky tethers have a stronger preference to align along the edges and vertices than shorter tethers. Even in the presence of moderate immiscibility, the conformational entropy dominates and bulky tethers lie along the edges and vertices. Intuitively, one might expect that vertices and edges are more preferable for bulky tethers than short tethers since the additional free volume will allow them to "fit better," increasing the overall system entropy. This is certainly true, but there is one surprising and noteworthy effect predicted by the theoretical model. At first glance, one might expect the disordered state to become stable in the $T \rightarrow \infty$ limit, but this is not the case: the ordered edge-aligned state remains stable at arbitrarily high temperatures. The reason is clear if one considers the mechanism behind the edge-alignment. In the $T \rightarrow \infty$ limit, any energetic effects leading to immiscibility at lower $T$ become negligible and the stable state is determined entirely by a competition between configurational and conformational entropy. If the entropy of the edge-aligned state is much larger than the entropy of the disordered state, then the system will remain ordered at arbitrarily high temperatures. This is the case when the difference between tether widths is large. It should be noted that the stability of the edge-aligned states at arbitrarily high temperatures will only be true for tethers made of hard spheres. For soft spheres like the ones used in the beads of the explicit tether model, the disordered state should become stable as $T \rightarrow \infty$ because the energetic interaction between tethers becomes negligible, allowing beads to overlap and resulting in a large free volume for all sites. Since real ligands interact with a potential that lies between soft and hard and our results show that both these potentials produce edge-aligned states, we expect the edge-aligned states to be stable in real systems for some finite range in temperature.

Although not explicitly stated, the free volume approximation relies on the assumption that the tether length is smaller than the particle size. If the ligands are considerably longer than the particle, the geometry of edges and faces would become less important because tethers in the middle of a face would be able to spill over the edges and the Voronoi cell approximation would break down. For this reason, there is likely an upper limit on the ratio of tether length to particle size. This is not necessarily an unwelcome restriction, since there are circumstances in which having short tethers and large particles is advantageous. For instance, this requirement forces the particle-tether system to maintain a faceted appearance even after tethers are added, which may be useful for assembly purposes described in the introduction. In addition, fluctuations in the concentration of tethers on the surface will be smaller for larger particles. However, this restriction remains an important consideration for experimentalists, since some applications involving patchy particles will depend on their size. Certain applications [57,58] require small domain sizes, which may necessitate the use of small particles.

## 4. Conclusions

Using a simple theoretical model in conjunction with simulations, we have predicted a novel entropically-ordered phase for mixed ligands adsorbed to a facetted nanoparticle surface. We found that sterically bulky ligands preferentially align along the vertices and edges of a faceted nanoparticle because this alignment maximizes the conformational entropy. The total entropy of the system is increased in the edge-aligned state because additional free volume available at vertices and edges permits a greater number of accessible configurations for bulky ligands bound at these sites than for those bound to a face site. We showed that this edge-aligned state is stable even in the presence of modest amounts of immiscibility. Our findings imply that entropic effects may provide a useful method of positioning ligands at precise locations on facetted nanoparticles. By controlling the location of chemically reactive ligands on particles, researchers might be able to design patchy particles that self-assemble into unique mesoscopic structures. It should be noted that this mechanism is general and in principle could be applied to a variety of polyhedrons provided the edges have sufficient curvature compared to the ligand architecture.

## 5. Acknowledgements

We would like to than Prof. Robert Ziff for his useful insights and suggestions for computing the entropy of a ligand. The authors acknowledge support from the James S. McDonnell Foundation 21st Century Science Research Award/Studying Complex Systems. The research was designed and conducted with support by the U.S. Department of Energy, Office of Basic Energy Sciences, Division of Materials Sciences and Engineering under Award DE-FG02-02ER46000. This material is based upon work supported by the DOD/DDRE under Award No. N00244-09-1-0062. Any opinions, findings, and conclusions or recommendations expressed in this publication are those of the author(s) and do not necessarily reflect the views of the DOD/DDRE.

## Notes and references

*a Department of Chemical Engineering, University of Michigan, 2300 Hayward St, Ann Arbor, MI 48109.*

*b 210 Wright Laboratory of Physics, Oberlin College, 110 North Professor Street, Oberlin, OH 44074.*

*c Department of Materials Science and Engineering, University of Michigan, 2300 Hayward St, Ann Arbor, MI 48109.*

*Corresponding author: sglotzer@umich.edu*